\pdfoutput=1

\documentclass[12pt,a4paper,twocolumn]{article}

\usepackage{ifthen} 
\usepackage{multicol}
\newboolean{pdflatex}
\setboolean{pdflatex}{true} 

\newboolean{articletitles}
\setboolean{articletitles}{true} 

\newboolean{uprightparticles}
\setboolean{uprightparticles}{false} 

\newboolean{inbibliography}
\setboolean{inbibliography}{false} 

\usepackage{microtype}
\usepackage{lineno}  
\usepackage{xspace} 
\usepackage{caption} 

\usepackage{graphicx}  
\usepackage{color}
\usepackage{colortbl}
\graphicspath{{./figs/}} 

\usepackage{amsmath} 
\usepackage{amssymb}
\usepackage{amsfonts}
\usepackage{upgreek} 

\newcommand*\patchAmsMathEnvironmentForLineno[1]{%
\expandafter\let\csname old#1\expandafter\endcsname\csname #1\endcsname
\expandafter\let\csname oldend#1\expandafter\endcsname\csname
end#1\endcsname
 \renewenvironment{#1}%
   {\linenomath\csname old#1\endcsname}%
   {\csname oldend#1\endcsname\endlinenomath}%
}
\newcommand*\patchBothAmsMathEnvironmentsForLineno[1]{%
  \patchAmsMathEnvironmentForLineno{#1}%
  \patchAmsMathEnvironmentForLineno{#1*}%
}
\AtBeginDocument{%
\patchBothAmsMathEnvironmentsForLineno{equation}%
\patchBothAmsMathEnvironmentsForLineno{align}%
\patchBothAmsMathEnvironmentsForLineno{flalign}%
\patchBothAmsMathEnvironmentsForLineno{alignat}%
\patchBothAmsMathEnvironmentsForLineno{gather}%
\patchBothAmsMathEnvironmentsForLineno{multline}%
\patchBothAmsMathEnvironmentsForLineno{eqnarray}%
}

\usepackage{hyperref}    
\usepackage[all]{hypcap} 

\def\lhcb {\mbox{LHCb}\xspace}

\ifthenelse{\boolean{uprightparticles}}%
{

 \def\PDelta      {\ensuremath{\Delta}\xspace}                 
 \def\PXi      {\ensuremath{\Xi}\xspace}                 
 \def\PLambda      {\ensuremath{\Lambda}\xspace}                 
 \def\PSigma      {\ensuremath{\Sigma}\xspace}                 
 \def\POmega      {\ensuremath{\Omega}\xspace}                 
 \def\PUpsilon      {\ensuremath{\Upsilon}\xspace}                 
 
 \def\PB      {\ensuremath{\mathrm{B}}\xspace}                 
                  
 \def\PD      {\ensuremath{\mathrm{D}}\xspace}

 \def\PK      {\ensuremath{\mathrm{K}}\xspace}

 \def\Pb      {\ensuremath{\mathrm{b}}\xspace}                 
 \def\Pc      {\ensuremath{\mathrm{c}}\xspace}

 \def\Pi      {\ensuremath{\mathrm{i}}\xspace}

 \def\Ps      {\ensuremath{\mathrm{s}}\xspace}

}
{

 \mathchardef\PDelta="7101
 \mathchardef\PXi="7104
 \mathchardef\PLambda="7103
 \mathchardef\PSigma="7106
 \mathchardef\POmega="710A
 \mathchardef\PUpsilon="7107
                  
 \def\PB      {\ensuremath{B}\xspace}                 
                  
 \def\PD      {\ensuremath{D}\xspace}

 \def\PK      {\ensuremath{K}\xspace}

 \def\Pb      {\ensuremath{b}\xspace}                 
 \def\Pc      {\ensuremath{c}\xspace}

 \def\Pi      {\ensuremath{i}\xspace}

 \def\Ps      {\ensuremath{s}\xspace}

}

\makeatletter
\ifcase \@ptsize \relax
  \newcommand{\miniscule}{\@setfontsize\miniscule{4}{5}}
\or
  \newcommand{\miniscule}{\@setfontsize\miniscule{5}{6}}
\or
  \newcommand{\miniscule}{\@setfontsize\miniscule{5}{6}}
\fi
\makeatother

\DeclareRobustCommand{\optbar}[1]{\shortstack{{\miniscule (\rule[.5ex]{1.25em}{.18mm})}
  \\ [-.7ex] $#1$}}

\def\squark    {{\ensuremath{\Ps}}\xspace}

\def\cquark    {{\ensuremath{\Pc}}\xspace}

\def\bquark    {{\ensuremath{\Pb}}\xspace}
\def\bquarkbar {{\ensuremath{\overline \bquark}}\xspace}
\def\bbbar     {{\ensuremath{\bquark\bquarkbar}}\xspace}

  \def\Kbar    {{\kern 0.2em\overline{\kern -0.2em \PK}{}}\xspace}

\def\KorKbar    {\kern 0.18em\optbar{\kern -0.18em K}{}\xspace}

  \def\Dbar    {{\kern 0.2em\overline{\kern -0.2em \PD}{}}\xspace}

\def\DorDbar    {\kern 0.18em\optbar{\kern -0.18em D}{}\xspace}

\def\Dzb     {{\ensuremath{\Dbar{}^0}}\xspace}

\def\B       {{\ensuremath{\PB}}\xspace}
\def\Bbar    {{\ensuremath{\kern 0.18em\overline{\kern -0.18em \PB}{}}}\xspace}

\def\BorBbar    {\kern 0.18em\optbar{\kern -0.18em B}{}\xspace}
\def\Bz      {{\ensuremath{\B^0}}\xspace}

\def\Bu      {{\ensuremath{\B^+}}\xspace}

\def\Bp      {{\ensuremath{\Bu}}\xspace}

\def\Bd      {{\ensuremath{\B^0}}\xspace}
\def\Bs      {{\ensuremath{\B^0_\squark}}\xspace}

  \def\Y#1S{\ensuremath{\PUpsilon{(#1S)}}\xspace}

\def\Lbar        {{\ensuremath{\kern 0.1em\overline{\kern -0.1em\PLambda}}}\xspace}
\def\LorLbar    {\kern 0.18em\optbar{\kern -0.18em \PLambda}{}\xspace}

\def\to                 {\ensuremath{\rightarrow}\xspace}

\def\AT#1     {\ensuremath{A_{\mathrm{T}}^{#1}}\xspace}           

\def\C#1      {\ensuremath{\mathcal{C}_{#1}}\xspace}                       
\def\Cp#1     {\ensuremath{\mathcal{C}_{#1}^{'}}\xspace}                    
\def\Ceff#1   {\ensuremath{\mathcal{C}_{#1}^{\mathrm{(eff)}}}\xspace}        
\def\Cpeff#1  {\ensuremath{\mathcal{C}_{#1}^{'\mathrm{(eff)}}}\xspace}       
\def\Ope#1    {\ensuremath{\mathcal{O}_{#1}}\xspace}                       
\def\Opep#1   {\ensuremath{\mathcal{O}_{#1}^{'}}\xspace}                    

\newcommand{\tev}{\ifthenelse{\boolean{inbibliography}}{\ensuremath{~T\kern -0.05em eV}\xspace}{\ensuremath{\mathrm{\,Te\kern -0.1em V}}}\xspace}
\newcommand{\gev}{\ensuremath{\mathrm{\,Ge\kern -0.1em V}}\xspace}
\newcommand{\mev}{\ensuremath{\mathrm{\,Me\kern -0.1em V}}\xspace}
\newcommand{\kev}{\ensuremath{\mathrm{\,ke\kern -0.1em V}}\xspace}
\newcommand{\ev}{\ensuremath{\mathrm{\,e\kern -0.1em V}}\xspace}
\newcommand{\gevc}{\ensuremath{{\mathrm{\,Ge\kern -0.1em V\!/}c}}\xspace}
\newcommand{\mevc}{\ensuremath{{\mathrm{\,Me\kern -0.1em V\!/}c}}\xspace}
\newcommand{\gevcc}{\ensuremath{{\mathrm{\,Ge\kern -0.1em V\!/}c^2}}\xspace}
\newcommand{\gevgevcccc}{\ensuremath{{\mathrm{\,Ge\kern -0.1em V^2\!/}c^4}}\xspace}
\newcommand{\mevcc}{\ensuremath{{\mathrm{\,Me\kern -0.1em V\!/}c^2}}\xspace}

\def\mm   {\ensuremath{\rm \,mm}\xspace}

\def\invfb   {\ensuremath{\mbox{\,fb}^{-1}}\xspace}

\newcommand{\stat}{\ensuremath{\mathrm{\,(stat)}}\xspace}
\newcommand{\syst}{\ensuremath{\mathrm{\,(syst)}}\xspace}

\def\gsim{{~\raise.15em\hbox{$>$}\kern-.85em
          \lower.35em\hbox{$\sim$}~}\xspace}
\def\lsim{{~\raise.15em\hbox{$<$}\kern-.85em
          \lower.35em\hbox{$\sim$}~}\xspace}

\def\pt         {\mbox{$p_{\rm T}$}\xspace}
\def\et         {\mbox{$E_{\rm T}$}\xspace}

\def\rad{\ensuremath{\rm \,rad}\xspace}

\def\evtgen     {\mbox{\textsc{EvtGen}}\xspace}

\def\gauss      {\mbox{\textsc{Gauss}}\xspace}
\def\geant      {\mbox{\textsc{Geant4}}\xspace}

\def\photos     {\mbox{\textsc{Photos}}\xspace}

\def\pythia     {\mbox{\textsc{Pythia}}\xspace}

\def\tell1  {TELL1\xspace}
\def\ukl1   {UKL1\xspace}

\usepackage{cite} 
\usepackage{mciteplus}

\usepackage{longtable} 

\def \afc {\ensuremath{A_{\rm C}^{b\bar{b}}}\xspace}

\def \mbb {\ensuremath{M_{b\bar{b}}}\xspace}
\def \Dy {\ensuremath{\Delta y}\xspace}

\begin{document}

\renewcommand{\thefootnote}{\fnsymbol{footnote}}
\setcounter{footnote}{1}

\onecolumn

\begin{titlepage}
\pagenumbering{roman}

\vspace*{-1.5cm}
\centerline{\large EUROPEAN ORGANIZATION FOR NUCLEAR RESEARCH (CERN)}
\vspace*{1.5cm}
\hspace*{-0.5cm}
\begin{tabular*}{\linewidth}{lc@{\extracolsep{\fill}}r}
\ifthenelse{\boolean{pdflatex}}
{\vspace*{-2.7cm}\mbox{\!\!\!\includegraphics[width=.14\textwidth]{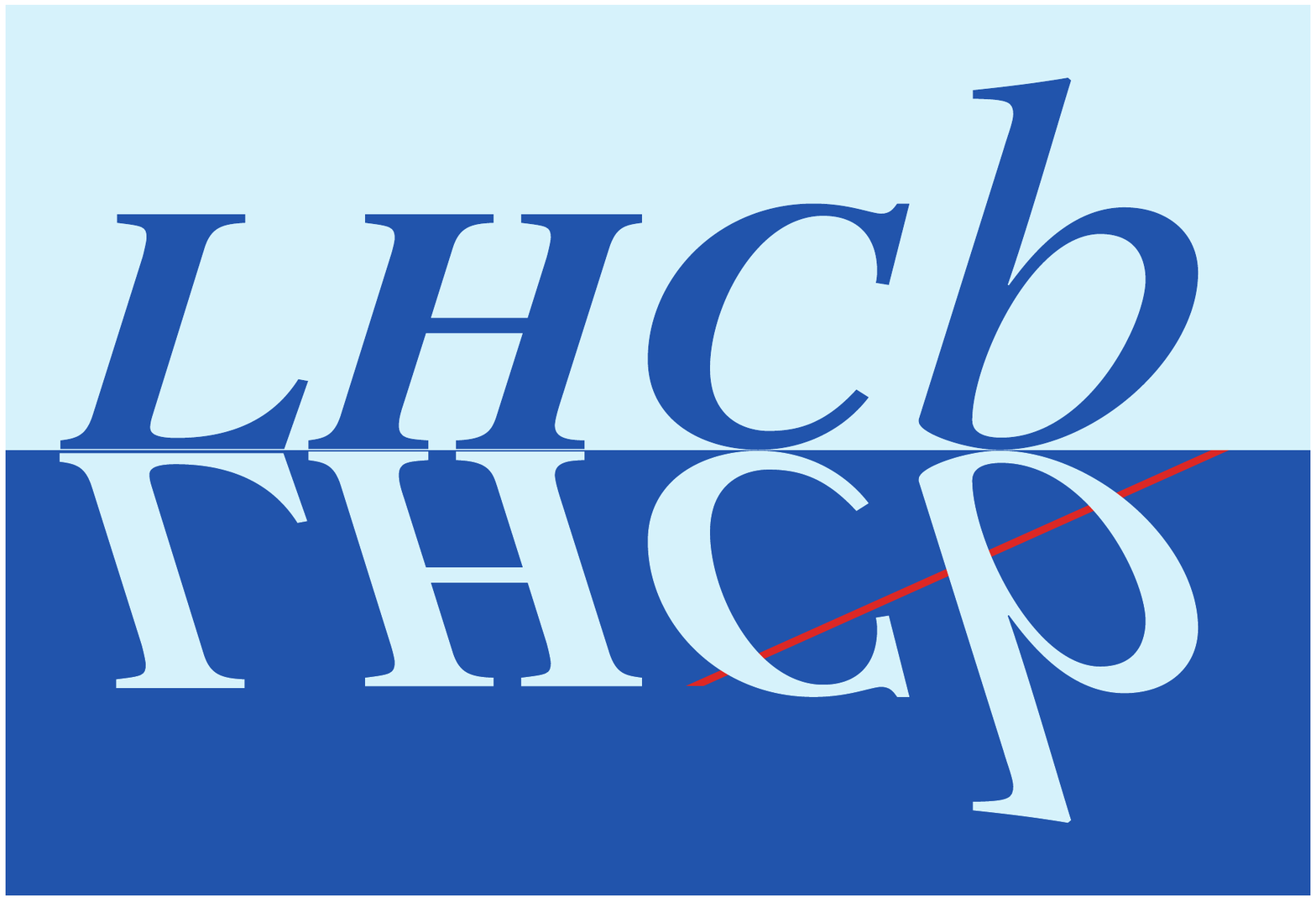}} & &}%
{\vspace*{-1.2cm}\mbox{\!\!\!\includegraphics[width=.12\textwidth]{lhcb-logo.eps}} & &}%
\\
 & & CERN-PH-EP-2014-130 \\  
 & & LHCb-PAPER-2014-023 \\  
 & & September 10, 2014 \\ 
 & & \\
\end{tabular*}

\vspace*{2.0cm}

{\bf\boldmath\huge
\begin{center}
  First measurement of the charge asymmetry in beauty-quark pair production at a hadron collider
\end{center}
}

\vspace*{0.0cm}

\begin{center}
The LHCb collaboration\footnote{Authors are listed on the following pages.}
\end{center}

\vspace{\fill}

\begin{abstract}
  \noindent
  The difference in the angular distributions between beauty quarks
  and antiquarks, referred to as the charge asymmetry, is measured for
  the first time in \bbbar pair production at a hadron collider.  The
  data used correspond to an integrated luminosity of 1.0\invfb
  collected at 7\tev center-of-mass energy in proton-proton collisions
  with the LHCb detector.  The measurement is performed in three
  regions of the invariant mass of the \bbbar system.  The results
  obtained are
  \begin{eqnarray}
    \afc (40 < \mbb < 75\gevcc) &=& 0.4 \pm 0.4\stat \pm 0.3\syst\%,  \nonumber\\
    \afc (75 < \mbb < 105\gevcc) &=& 2.0 \pm 0.9\stat \pm 0.6\syst\%,  \nonumber\\
    \afc (\mbb > 105\gevcc) &=& 1.6 \pm 1.7\stat \pm 0.6\syst\%, \nonumber
\end{eqnarray}
where \afc is defined as the asymmetry in the difference in rapidity between jets formed from the beauty quark and antiquark.  
The beauty jets are required to satisfy $2<\eta<4$, $\et > 20\gev$, and have an opening angle in the transverse plane $\Delta\phi>2.6$\,rad.
These measurements are consistent with the predictions of the Standard Model.
\end{abstract}

\vspace*{2.0cm}

\vspace{\fill}

{\footnotesize 
\centerline{\copyright~CERN on behalf of the \lhcb collaboration, license \href{http://creativecommons.org/licenses/by/3.0/}{CC-BY-3.0}.}}
\vspace*{2mm}

\end{titlepage}


\newpage
\setcounter{page}{2}
\mbox{~}
\newpage

\centerline{\large\bf LHCb collaboration}
\begin{flushleft}
\small
R.~Aaij$^{41}$, 
B.~Adeva$^{37}$, 
M.~Adinolfi$^{46}$, 
A.~Affolder$^{52}$, 
Z.~Ajaltouni$^{5}$, 
S.~Akar$^{6}$, 
J.~Albrecht$^{9}$, 
F.~Alessio$^{38}$, 
M.~Alexander$^{51}$, 
S.~Ali$^{41}$
% ...

{\footnotesize \it
% ...
$ ^{5}$Clermont Universit\'{e}, Universit\'{e} Blaise Pascal, CNRS/IN2P3, LPC, Clermont-Ferrand, France\\
$ ^{6}$CPPM, Aix-Marseille Universit\'{e}, CNRS/IN2P3, Marseille, France\\
% ...
$ ^{9}$Fakult\"{a}t Physik, Technische Universit\"{a}t Dortmund, Dortmund, Germany\\
% ...
$ ^{37}$Universidad de Santiago de Compostela, Santiago de Compostela, Spain\\
$ ^{38}$European Organization for Nuclear Research (CERN), Geneva, Switzerland\\
% ...
$ ^{41}$Nikhef National Institute for Subatomic Physics, Amsterdam, The Netherlands\\
% ...
$ ^{46}$H.H. Wills Physics Laboratory, University of Bristol, Bristol, United Kingdom\\
% ...
$ ^{51}$School of Physics and Astronomy, University of Glasgow, Glasgow, United Kingdom\\
$ ^{52}$Oliver Lodge Laboratory, University of Liverpool, Liverpool, United Kingdom\\
% ...
}
\end{flushleft}

\cleardoublepage

\twocolumn

\renewcommand{\thefootnote}{\arabic{footnote}}
\setcounter{footnote}{0}


\pagestyle{plain} 
\setcounter{page}{1}
\pagenumbering{arabic}


Measurements 
in $p\bar{p}$ collisions
at the Tevatron~\cite{D0ttbarOld,D0diLep,D0ljets,CDFttbar,CDFljets,CDFdiLep}
suggest that (anti)top quarks are produced along the (anti)proton beam
direction more often than predicted by the Standard Model
(SM)~\cite{benreuther}.  Many extensions to the SM have been proposed
to explain this discrepancy (for a review, see Ref.~\cite{saavedra}) that couple new particles to
quarks in a variety of ways. Therefore, constraints on quark-antiquark
production charge asymmetries other than top-anti-top ($t\bar{t}$)
could discriminate between models and be used as a probe of non-SM
physics.  For example, some theories proposed to explain the Tevatron
results also predict a large charge asymmetry in $b\bar{b}$
production~\cite{afc,afc2}.  No measurement has been made to date of the
\bbbar charge asymmetry at a hadron collider.

The symmetric initial state of proton-proton collisions at the LHC
does not permit a charge asymmetry to be manifest as an observable
defined using the direction of one beam relative to the
other. However, the asymmetry in the momentum fraction of quarks and
antiquarks inside the proton means that a charge asymmetry can lead to
a difference in the rapidity distributions of beauty quarks and
antiquarks.  The $b\bar{b}$ charge asymmetry in $pp$ collisions is
defined as 
\begin{equation}
  \label{eq:afc}
  \afc \equiv \frac{N(\Delta y > 0) - N(\Delta y < 0)}{N(\Delta y > 0) + N(\Delta y < 0)},
\end{equation}
where $\Dy \equiv |y_{b}| - |y_{\bar{b}}|$ is the rapidity difference
between jets formed from the $b$ and $\bar{b}$ quarks.  Measurements
of the top-quark charge asymmetry by the ATLAS and CMS experiments are
consistent with the SM expectations~\cite{afctt1,afctt2,afctt3}.
However, the large $gg\to t\bar{t}$ cross-section at the LHC dilutes
the observable signal of new physics entering the $q\bar{q}\to
t\bar{t}$ process that dominates $t\bar{t}$ production at the
Tevatron.

In the SM the only sizable leading-order (LO) contribution to \afc
comes from $Z \to \bbbar$ decays.  The contribution of $Z\to\bbbar$ to
\afc in a region of invariant mass of the \bbbar system (\mbb) around
the $Z$ boson mass is expected to be about 2\% based on simulation.
Production of \bbbar pairs at LO in
quantum chromodynamics (QCD) is symmetric under the exchange of $b$
and $\bar{b}$ quarks. At higher orders, radiative corrections to the
$q\bar{q}\to\bbbar$ process generate an asymmetry in the differential
distributions of the $b$ and $\bar{b}$ quarks and induce a correlation
between the direction of the $b$ ($\bar{b}$) quark and that of the
incoming $q$ ($\bar{q}$) quark.  Such higher-order corrections are
expected to be negligible at low \mbb
and to increase in
importance at larger \mbb.  The contribution to \afc from higher-order
terms is expected to reach 1\% near the $Z$ boson mass~\cite{gauld}.
Precision measurements of \afc as a function of \mbb are sensitive probes of
physics beyond the SM.

This Letter reports the first measurement of the charge asymmetry in
beauty-quark pair production at a hadron collider.  The data used
correspond to an integrated luminosity of 1.0\invfb collected at 7\tev
center-of-mass energy in $pp$ collisions with the \lhcb detector.  The
measurement is performed in three regions of \mbb: ${40 < \mbb <
  75\gevcc}$, ${75 < \mbb < 105\gevcc}$ and ${\mbb > 105\gevcc}$.
This scheme is chosen such that the middle region is centered around
the mass of the $Z$ boson and contains most of the $Z \to \bbbar$
candidates.
The measurement is corrected to a pair
of particle-level jets, each with a pseudorapidity $2<\eta<4$,
transverse energy $\et > 20\gev$, and an opening angle between the
jets in the transverse plane $\Delta\phi>2.6$\,rad.

The \lhcb detector is a single-arm forward spectrometer covering the
range $2<\eta <5$ designed for the study of particles containing
\bquark or \cquark quarks, described in detail in
Refs.\cite{Alves:2008zz,LHCb-DP-2013-003,LHCb-DP-2012-003,LHCb-DP-2012-002}.
The trigger~\cite{LHCb-DP-2012-004} consists of a hardware stage,
based on information from the calorimeter and muon systems, followed
by a software stage, which applies a full event reconstruction.
Identification of beauty-hadron decays in the software trigger
requires a two-, three- or four-track secondary vertex with a large
sum of the transverse momentum (\pt) of the tracks and a significant
displacement from the primary $pp$ interaction vertices.  A
multivariate algorithm~\cite{BBDT} is used for the identification of
vertices consistent with the decay of a beauty hadron.  This so-called
topological trigger algorithm (TOPO) is also used in this analysis to
identify the hadrons that contain the beauty quark and antiquark in
\bbbar pair production.  The charge of the beauty (anti)quarks is
determined by the charge of muons originating from semileptonic
beauty-hadron decays.

Simulated events are used to calibrate the jet energy scale, to
determine the reconstruction and selection efficiencies and to unfold
the detector response.  In the simulation, $pp$ collisions are
generated using \pythia~\cite{Sjostrand:2006za} with a specific \lhcb
configuration~\cite{LHCb-PROC-2010-056}.  Decays of hadronic particles
are described by \evtgen~\cite{Lange:2001uf}, in which final state
radiation is generated using \photos~\cite{Golonka:2005pn}. The
interaction of particles with the detector and its
response are implemented using the \geant
toolkit~\cite{Allison:2006ve, *Agostinelli:2002hh} as described in
Ref.~\cite{LHCb-PROC-2011-006}.

The \bbbar are reconstructed as jets using the
anti-$k_{\rm T}$ algorithm~\cite{1126-6708-2008-04-063} with distance
parameter $R=0.7$, as implemented in
\textsc{Fastjet}~\cite{Cacciari200657}. The inputs to the jet
reconstruction are selected using a particle flow
approach~\cite{LHCb-PAPER-2013-058}. Information from all the detector
sub-systems is used to create charged and neutral particle inputs to
the jet algorithm. 
Jet-quality criteria are applied to
remove jets for which a large fraction of the energy is likely due to
sources other than a $pp$ collision, {\em e.g.}, detector noise or
poorly reconstructed tracks. The per-jet efficiency of these criteria
is $90-95\%$ depending on the jet kinematic properties.  The mean
number of $pp$ collisions per event 
is only 1.8 making it unlikely to produce \bbbar in separate collisions; however, to prevent this, both jets are required to originate from the same $pp$
collision.

The observed energy of each jet is corrected to the particle-level
energy accounting for the following effects: imperfect detector
response; the presence of detector noise; energy contributions from
$pp$ interactions other than the one in which the \bbbar are produced; beauty (anti)quark energy flowing out of the
jet cone; and the presence of a neutrino from the semileptonic decay
of a beauty hadron in the jet.  The jet energy correction varies in
the range $0-20\%(\pm10\%)$ for jets that do(do not) contain a
neutrino from a semileptonic beauty-hadron decay.  The mean value for
jets that do not contain a semileptonic-decay neutrino is about 1\%.
This correction is obtained from simulation and depends on the jet
$\eta$, \et, and the number of $pp$ interactions in the event.  Only
jets in a well-understood kinematic regime of \lhcb, $\et > 20\gev$
and $2 < \eta < 4$, are considered in this analysis.  The relative
resolution on \mbb obtained using these jets is about 15\%.

Jets in events selected by the TOPO need to be identified (tagged) as
containing a beauty quark or antiquark (bTAG).  For this task, an
association is made between jets and the multitrack TOPO objects.  If
at least 60\% of the detector hits that make up the tracks forming the
TOPO object also belong to tracks within the jet, then the jet
satisfies a bTAG requirement.  At least one jet in the event is
required to contain a beauty-hadron decay selected by the TOPO which
caused the event to be recorded.  The TOPO is applied to
offline-reconstructed tracks with a looser requirement to search for a
second beauty-hadron decay in the event.  If such a decay is found,
and if it can be associated to another jet, then the event is
identified as containing a \bbbar pair. The mean di-bTAG efficiency
for di-jet events used in this analysis is about 30\%, while the
per-jet mis-tag efficiency for jets initiated by light quarks and
gluons is less than 0.1\%.  To enhance the contribution of non-$gg$
production mechanisms, $\Delta\phi>2.6~\rad$ is required between the
two jets that satisfy the bTAG requirement.

The largest background contribution is due to charm jets. 
The level of background
contamination is determined using the so-called corrected mass
\begin{equation}
M_{\rm corr} = \sqrt{M^2 + \left (\frac{p}{c}\right )^2\sin^2{\theta}} + \frac{p}{c}\sin{\theta},
\end{equation}
where $M$ and $p$ are the invariant mass and momentum of all tracks in
the jet that are inconsistent with originating directly from a $pp$
collision and have a minimum distance of closest approach to a track
used in the TOPO less than 0.2\mm.  The angle $\theta$ is between the
momentum and the direction from the $pp$ collision to the TOPO object
vertex.  The corrected mass is the minimum mass the long-lived hadron
can have that is consistent with the direction of flight.

\begin{figure}[] 
\centering
\includegraphics[width=0.49\textwidth]{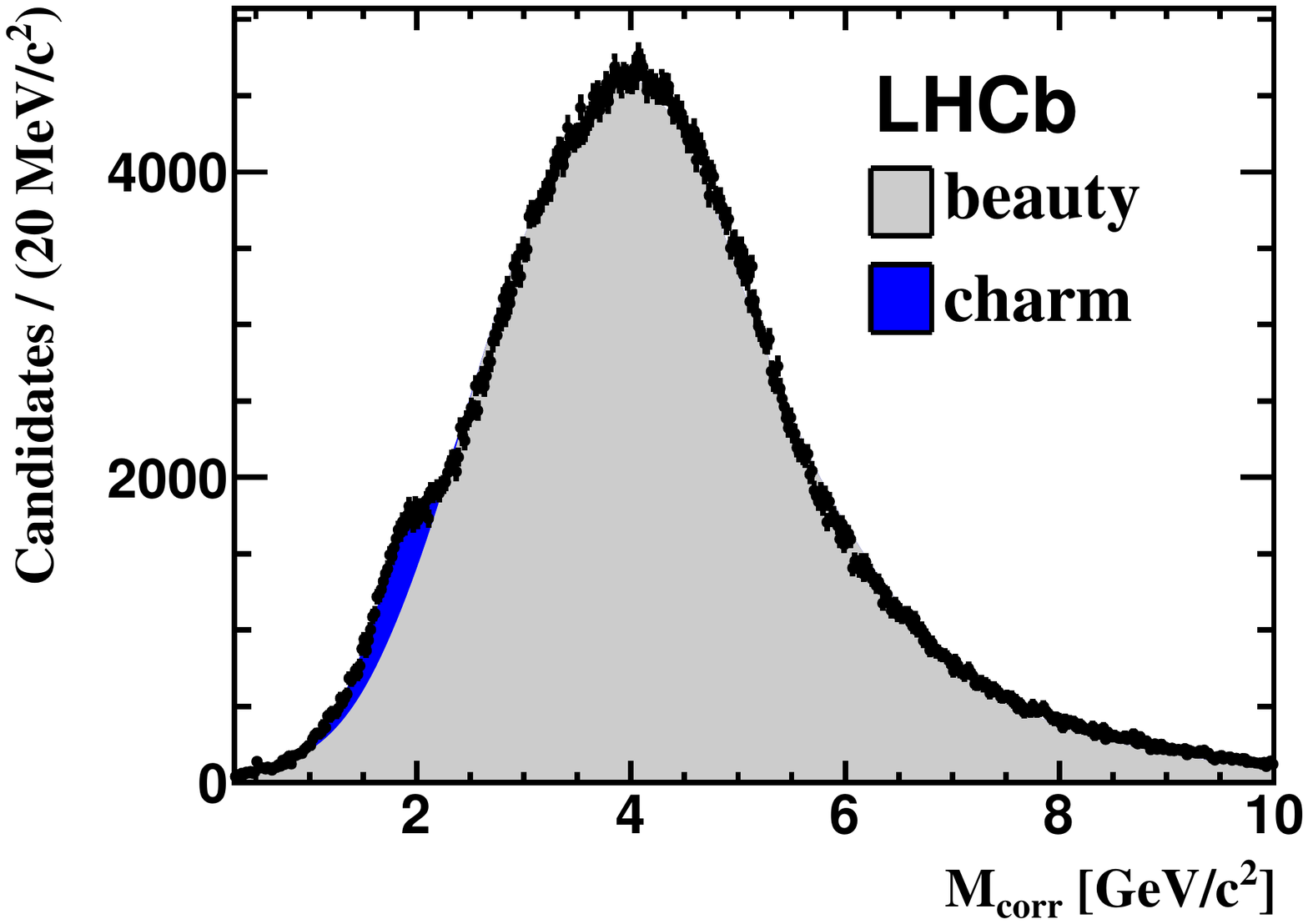}
\includegraphics[width=0.49\textwidth]{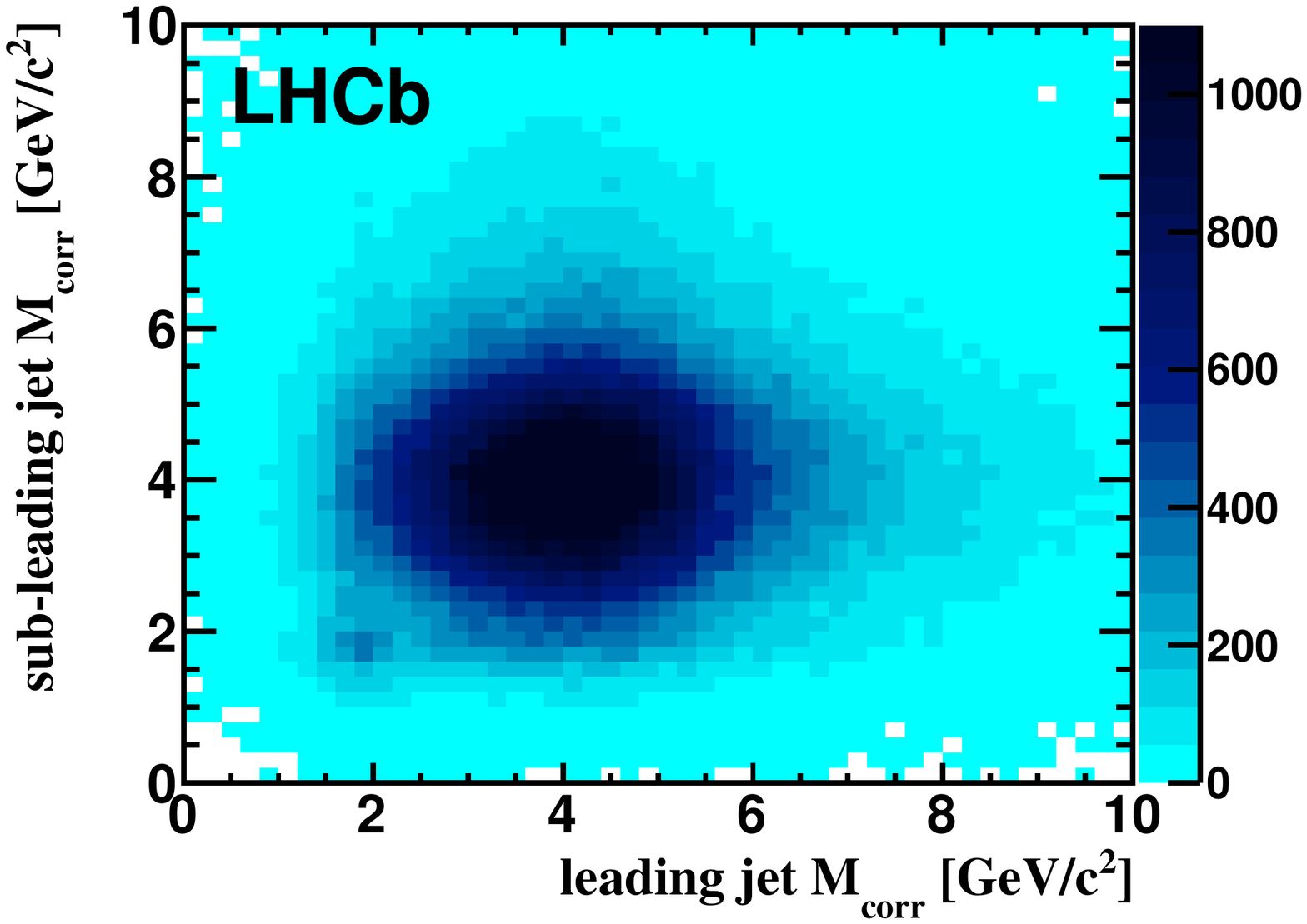}
\caption{(top) Corrected mass of TOPO objects associated to bTAG jets in the final event sample.  Less than 2\% of jets are found to originate from charm.  (bottom) Corrected mass of TOPO objects associated to sub-leading {\em vs} leading jets in the final event sample.  A small $c\bar{c}$ contribution is visible near (2,2)\gevcc.}
\label{fig:mcor}
\end{figure}

Figure~\ref{fig:mcor} shows the corrected-mass distribution. 
The corrected-mass probability density functions (PDFs) for beauty and
charm are obtained from simulation.  Imperfect measurement of the
direction of flight can result in a larger corrected mass than the
true hadron mass.  For charm-hadron decays, the particles originate
from a single point in space and typically the missing momentum is
carried by a single low-mass particle; thus, the corrected mass peaks
near the known charm-meson mass.  
The vast majority of beauty-hadron decays involve intermediate charm hadrons which results in not all stable particles originating from the same spatial point.
The missing momentum is typically carried away by multiple particles
and the invariant mass of the missing momentum may be large.  Hence,
the corrected mass for beauty decays peaks below the known
beauty-meson mass and has worse resolution than for charm.
The result of a fit to the data shown in
Fig.~\ref{fig:mcor} is that $3.6\pm1.2\%$ of events in the final
sample are not \bbbar, where the uncertainty is due to the corrected-mass PDFs.  The contribution from jets initiated by
light quarks or gluons is found to be negligible. Furthermore, the
limited acceptance of the LHCb detector for $b\bar{b}$ originating
from $t\bar{t}$ makes this contribution negligible as well.

To measure the charge asymmetry, the charge of the beauty (anti)quark
needs to be identified in at least one of the jets (qTAG).  The qTAG
requirement is that a track in the TOPO object and in the jet is
identified as a muon.  The muon is required to satisfy $\pt > 2\gevc$
and $p > 10\gevc$ to reduce the charge asymmetry due to detector
biases.  This strategy is designed to look for muons coming from
semileptonic beauty-hadron decays; thus, the charge of the muon tags
the charge of the beauty quark or antiquark.  Decays of the type $b
\to c \to \mu$ contaminate the charge tagging.  To mitigate this, the
tagging muon is required to have the highest momentum of all displaced
tracks in the jet.  A further dilution to the charge-tagging purity
arises due to oscillations of the \Bz and \Bs mesons.

The expected qTAG purity, defined as the probability to correctly
assign the charge of the beauty quark in a qTAG jet, can be estimated
using the following: the measured $b$-hadron production
fractions~\cite{LHCb-CONF-2013-011,LHCb-PAPER-2011-018}; the
$b$-hadron and $c$-hadron semileptonic branching
fractions~\cite{PDG2012}; the charge-tagging efficiencies for $b$ and
$c$-hadron semileptonic decays obtained from simulation; the \Bd and
\Bs oscillation
frequencies~\cite{LHCb-PAPER-2013-006,LHCb-PAPER-2012-032} and the
reconstruction efficiency as a function of $b$-hadron lifetime
obtained from simulation.  Combining all of this information yields an
expected qTAG purity of $73\pm4\%$.  The purity is expected to
decrease by a few percent with increasing jet energy due to 
an increase in the neutral-beauty-meson production fractions.

The qTAG purity is measured directly using events where both bTAG jets
also satisfy the qTAG requirement using the fraction of events where
the two muons have opposite charges.  This gives an integrated qTAG
purity of $70.3\pm0.3\%$, which agrees with the predicted value, and
values of $71.6\pm0.5\%$, $68.8\pm0.8\%$ and $66.1\pm1.9\%$ for $40 <
\mbb < 75\gevcc$, $75 < \mbb < 105\gevcc$ and $\mbb > 105\gevcc$,
respectively.  The observed decrease in purity agrees with
expectations.
The qTAG purity is found to be consistent in data for all \Dy.  
As a further consistency check, a separate study of the qTAG purity is performed
using events with a jet and a fully reconstructed self-tagging $B^+
\to J/\psi K^+$ or $B^+ \to \Dzb \pi^+$ decay.  In these events, the
charge of the \Bp provides an unambiguous qTAG of the beauty jet for
\bbbar pair production.  Using $B^+ +$jet events where the jet
satisfies the qTAG, the qTAG purity is determined to be $73\pm3\%$.
This result agrees with both the predicted and di-qTAG results. The
di-qTAG purity measurement is used to obtain the final \afc
results below.

\begin{figure}[t] 
\centering
\includegraphics[width=0.49\textwidth]{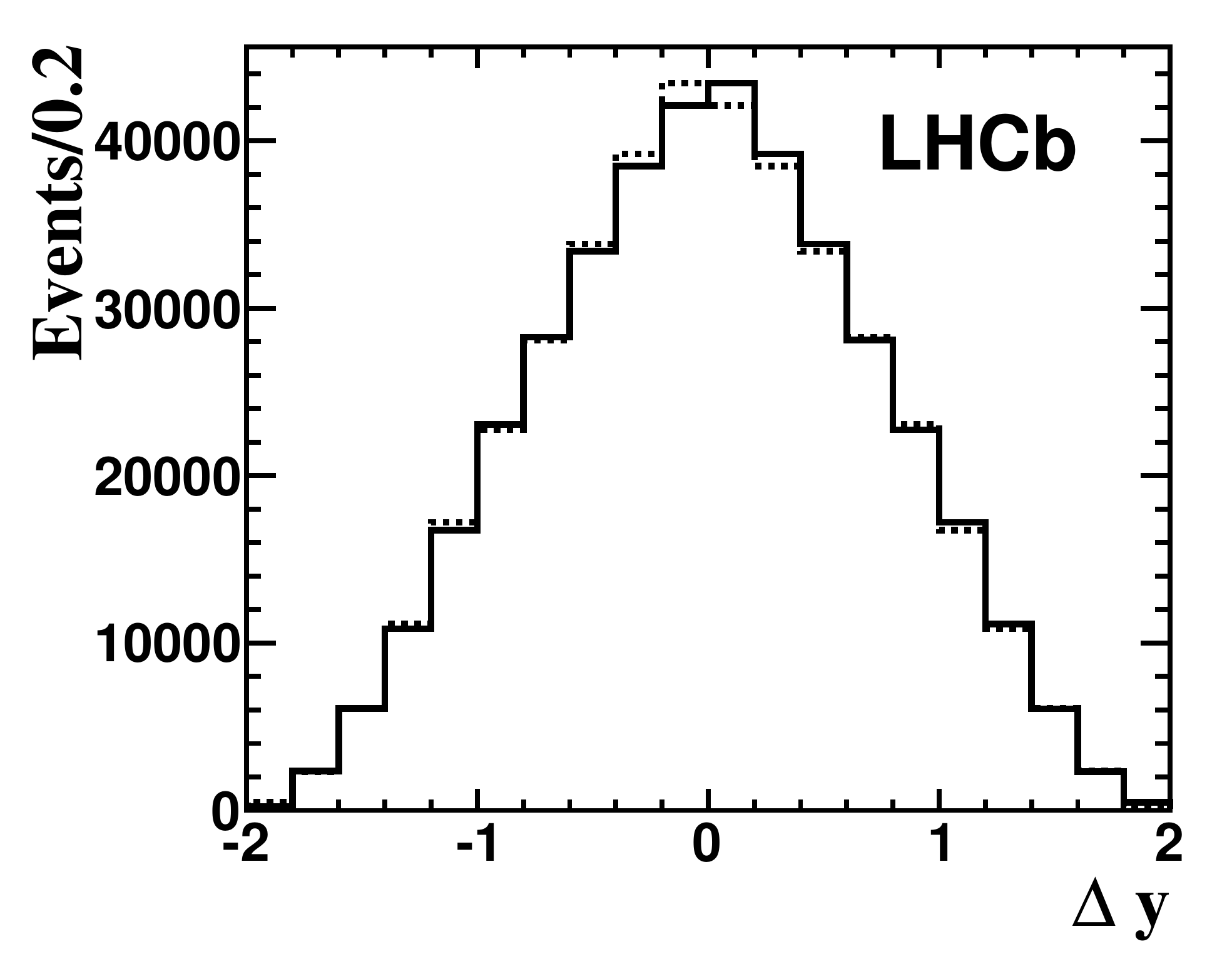}
\caption{Reconstructed $\Delta y$ distribution for all selected events
  after background subtraction and correction for qTAG impurity.  The
  dashed line shows the distribution reflected about the vertical axis.}
\label{fig:dycor}
\end{figure}

\begin{figure}[] 
\centering
\includegraphics[width=0.5\textwidth]{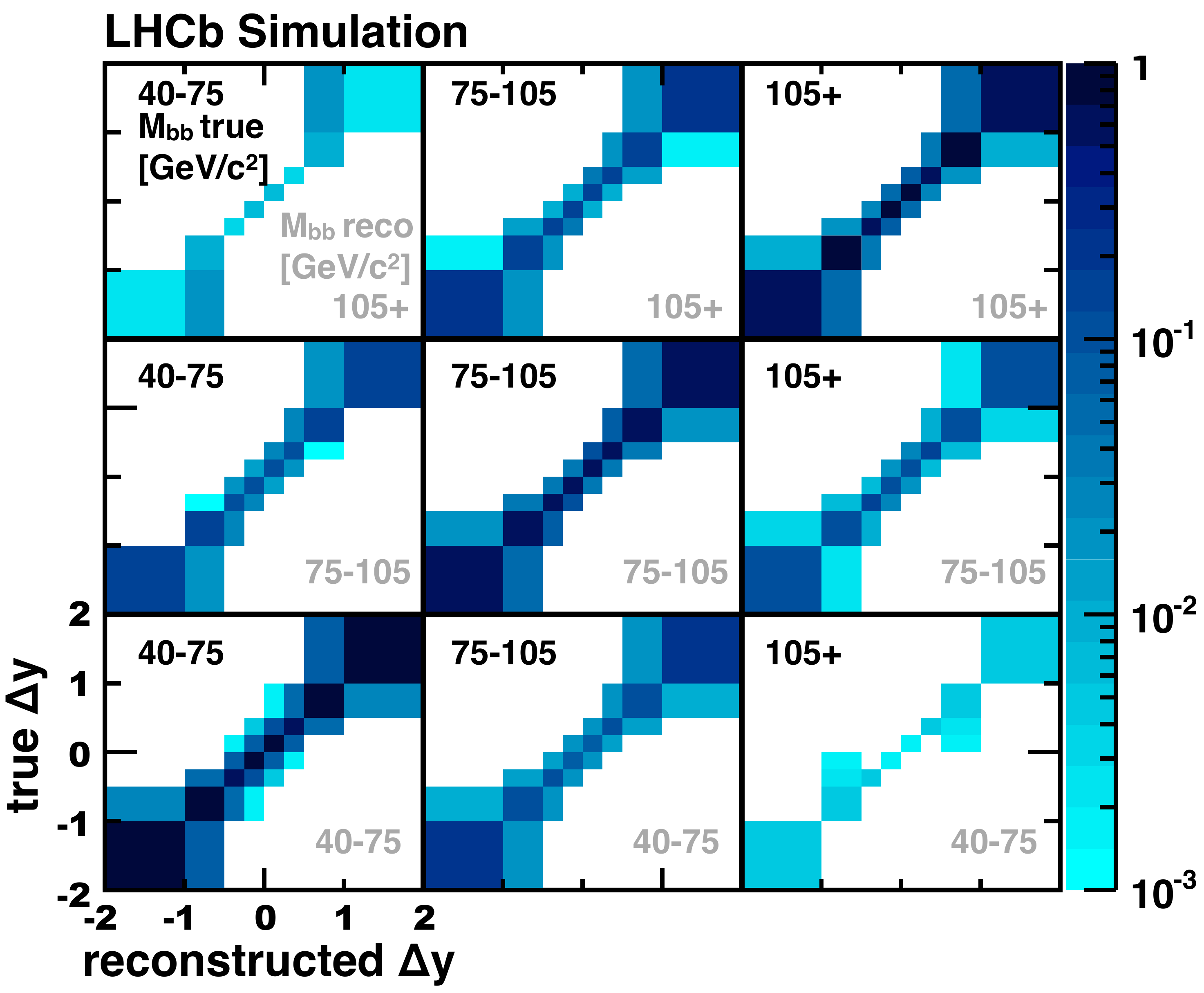}
\caption{Migration matrix in \Dy and \mbb.}
\label{fig:response}
\end{figure}

Figure~\ref{fig:dycor} shows the \Dy distribution 
after background subtraction and
correcting for qTAG impurity.  
The reconstructed distributions of \Dy and
\mbb are corrected for the effects of detector resolution and for
event reconstruction and selection efficiency.  The correction for
detector resolution is achieved by applying a two-dimensional
unfolding procedure to the data~\cite{Agostini}. The migration matrix
in \Dy and \mbb is shown in Fig.~\ref{fig:response}.  
The selection efficiency is obtained from
simulated events as a function of \Dy and \mbb. The residual
dependence of the efficiency on other jet kinematic variables has a
negligible impact on the resulting measurement of \afc.

The main sources of systematic uncertainties on the measurement of
\afc are as follows: precision of the qTAG purity and its dependence
on jet kinematic properties; uncertainty in the unfolding;
determination of the selection efficiency; and any residual
detector-related asymmetries.  Table~\ref{tab:afc_sys} summarizes the
values of the systematic uncertainties assigned to the measurement of
\afc in each \mbb region.  Measurement of the qTAG purity is
data-driven and the statistical uncertainties are propagated to \afc
to determine the systematic uncertainty.  The uncertainty due to
unfolding accounts for the choice of data sample used to generate the
migration matrix and mismodeling of the detector response in the
simulation.  
The uncertainty due to efficiency is dominated by the statistical
uncertainty of the simulation.  The polarity of the LHCb dipole magnet
is reversed periodically. This coupled with the hard momentum spectrum
of the tagging muons results in only small detection-based
asymmetries. Additionally, due to the definition of \Dy, these
detection asymmetries cancel to very good approximation when summing
over $\mu^+$ and $\mu^-$ tags. The detection asymmetry of charged
kaons causes a negligible bias in \afc.  

\begin{figure}[t] 
\centering
\includegraphics[width=0.5\textwidth]{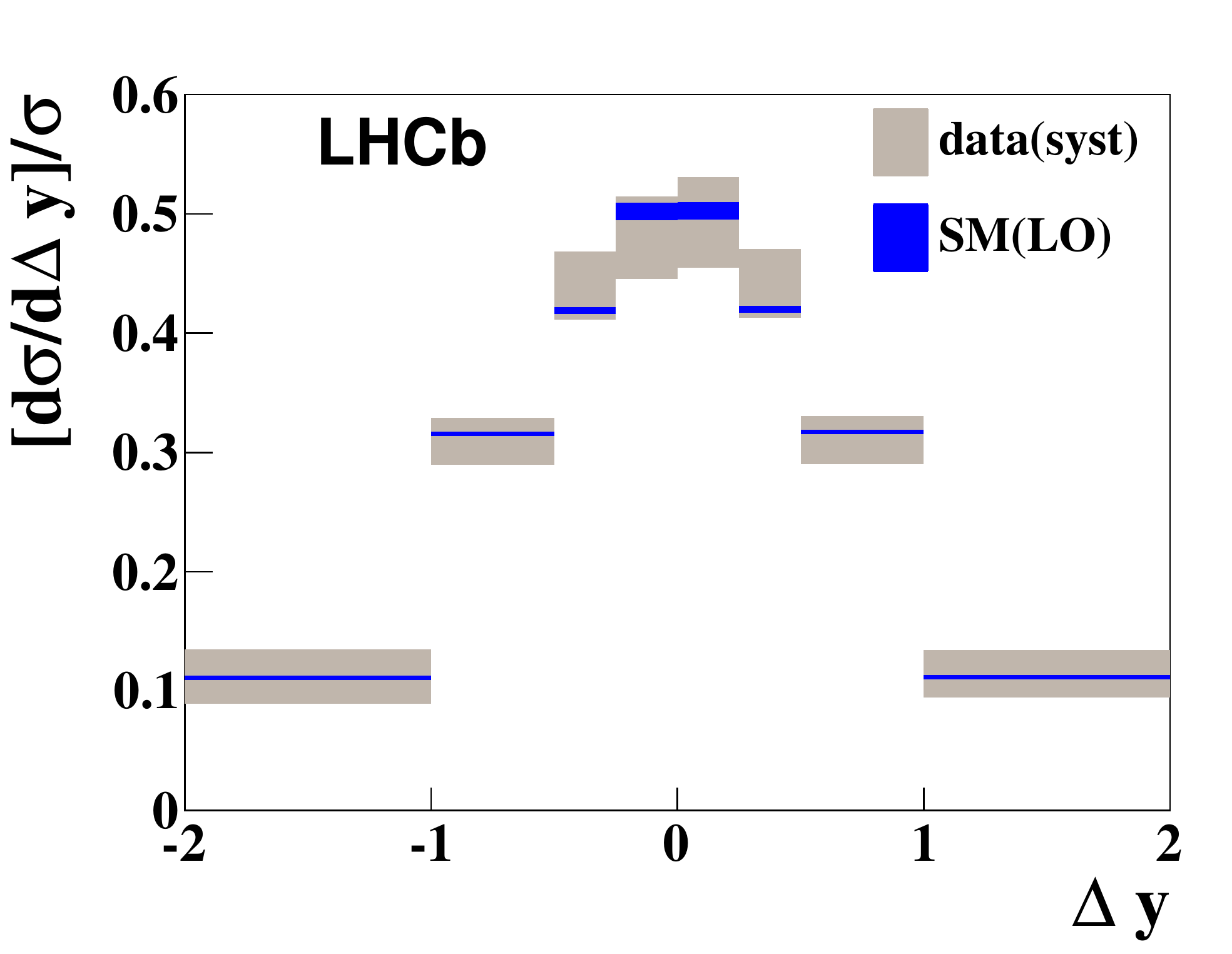}
\caption{Corrected \Dy distribution for all selected events.  The
  statistical uncertainties are negligible.  The systematic
  uncertainties are highly correlated from bin to bin and largely
  cancel in the determination of \afc.  The LO SM prediction obtained
  from \pythia~\cite{Sjostrand:2007gs,cteq6l1} is also shown.  }
\label{fig:inclDy}
\end{figure}

Figure~\ref{fig:inclDy} shows the corrected $\Dy$ distribution summed
over all \mbb regions considered ($\mbb > 40\gevcc$).  The LO SM
prediction, which includes LO QCD and $Z\to b\bar{b}$, obtained from
\pythia~\cite{Sjostrand:2007gs,cteq6l1} is also shown.  The SM uncertainty
includes contributions from the renormalization and factorization
scales, and from the parton distribution functions.  A next-to-LO SM
calculation is required to obtain \afc at the percent level.  However,
the LO result is sufficient to demonstrate agreement between the
theory and unfolded \bbbar pair-production distribution.

The measurement of \afc is performed in three regions of \mbb and the results obtained are
\begin{eqnarray}
  \afc (40,75)   &\hspace{-0.5em}=&\hspace{-0.5em} 0.4 \pm 0.4\stat \pm 0.3\syst\%,  \nonumber\\
  \afc (75,105) &\hspace{-0.5em}=&\hspace{-0.5em} 2.0 \pm 0.9\stat \pm 0.6\syst\%,  \nonumber\\ 
  \afc (>105) &\hspace{-0.5em}=&\hspace{-0.5em} 1.6 \pm 1.7\stat \pm 0.6\syst\%, \nonumber
\end{eqnarray}
where the ranges denote the regions of \mbb in units of \gevcc.
These measurements are the first to date of the charge asymmetry in
\bbbar pair production at a hadron collider.  The results are
corrected to a pair of particle-level jets each with $2<\eta<4$, $\et
> 20\gev$ and $\Delta\phi>2.6$\,rad between the jets. All results are
consistent with the SM expectations.

\begin{table}
  \centering
  \caption{Absolute systematic uncertainties (\%).} 
  \begin{tabular}{cccc}
    \hline\\[-1.85ex]
    {} & \multicolumn{3}{c}{\mbb (GeV/$c^2$)} \\
    Source & $(40,75)$ & $(75,105)$ & $>105$ \\ \hline
    Mis-qTAG & $-$ & 0.1 & 0.2 \\
    Unfolding  & 0.3 & 0.6 & 0.4 \\
    $\varepsilon(\mbb,\Dy)$ & 0.1 & 0.1 & 0.2 \\
    $\varepsilon(b)-\varepsilon(\bar{b})$ & $-$ & $-$ & 0.3\\
    \hline
    Total & 0.3 & 0.6 & 0.6 \\
    \hline
  \end{tabular}
  \label{tab:afc_sys}
\end{table}

\section*{Acknowledgements}
 \noindent We express our gratitude to our colleagues in the CERN
accelerator departments for the excellent performance of the LHC. We
thank the technical and administrative staff at the LHCb
institutes. We acknowledge support from CERN and from the national
agencies: CAPES, CNPq, FAPERJ and FINEP (Brazil); NSFC (China);
CNRS/IN2P3 (France); BMBF, DFG, HGF and MPG (Germany); SFI (Ireland);
INFN (Italy); FOM and NWO (The Netherlands); MNiSW and NCN (Poland);
MEN/IFA (Romania); MinES and FANO (Russia); MinECo (Spain); SNSF and
SER (Switzerland); NASU (Ukraine); STFC (United Kingdom); NSF (USA).
The Tier1 computing centres are supported by IN2P3 (France), KIT and
BMBF (Germany), INFN (Italy), NWO and SURF (The Netherlands), PIC
(Spain), GridPP (United Kingdom).  We are indebted to the communities
behind the multiple open source software packages on which we
depend. We are also thankful for the computing resources and the
access to software R\&D tools provided by Yandex LLC (Russia).
Individual groups or members have received support from EPLANET, Marie
Sk\l{}odowska-Curie Actions and ERC (European Union), Conseil
g\'{e}n\'{e}ral de Haute-Savoie, Labex ENIGMASS and OCEVU, R\'{e}gion
Auvergne (France), RFBR (Russia), XuntaGal and GENCAT (Spain), Royal
Society and Royal Commission for the Exhibition of 1851 (United
Kingdom).

\addcontentsline{toc}{section*}{References}
\setboolean{inbibliography}{true}
\ifx\mcitethebibliography\mciteundefinedmacro
\PackageError{LHCb.bst}{mciteplus.sty has not been loaded}
{This bibstyle requires the use of the mciteplus package.}\fi
\providecommand{\href}[2]{#2}

\onecolumn
\end{document}